# A THEORETICAL FRAMEWORK FOR RETINAL COMPUTATIONS:

# INSIGHTS FROM TEXTBOOK KNOWLEDGE

SAMUEL CHIQUITA

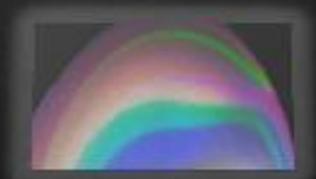

"Some secret truths, from learned pride concealed,

To maids alone and children are revealed:

What though no credit doubting wits may give?

The fair and innocent shall still believe."

Alexander Pope *in The Rape of the Lock*

# A theoretical framework for retinal computations: insights from textbook knowledge

Samuel Chiquita[1]

[1]University of Porto, Porto, Portugal

**Abstract**

Neural circuits in the retina divide the incoming visual scene into more than a dozen distinct representations that are sent on to central brain areas, such as the lateral geniculate nucleus and the superior colliculus. The retina can be viewed as a parallel image processor made of a multitude of small computational devices. Neural circuits of the retina are constituted by various cell types that separates the incoming visual information in different channels. Visual information is processed by retinal neural circuits and several computations are performed extracting distinct features from the visual scene. The aim of this article is to understand the computational basis involved in processing visual information which finally leads to several feature detectors. Therefore, the elements that form the basis of retinal computations will be explored by explaining how oscillators can lead to a final output with computational meaning. Linear versus nonlinear systems will be presented and the retina will be placed in the context of a nonlinear system. Finally, simulations will be presented exploring the concept of the retina as a nonlinear system which can perform understandable computations converting a known input into a predictable output.

**Contents**

# 1. Introduction

"The book of nature is written in the language of mathematics", once Galileo stated [1]. The possibility to describe the phenomena that exist in the universe with mathematics is an appealing way of looking to reality, a possibility through which reality can be more tangible and understandable. The processes that happen in life, the processes that makes us Human are of particular interest. Specifically, Neuroscience studies the processes that happen at the neural level and that can make us aware of ourselves and others. In these processes, which might be called cognition or emotions, information is somehow encoded on the neural activity. This encoding process probably is present at the level of the action potentials spike trains, which can be a synchronous symphony present in our own biological system. This encoding might be called computation. How is then information registered in our body? A possible way to address this question is to focus attention into human special senses, such as vision. In this system the information from the visual scene is integrated firstly at the retina and then projects to other brain regions, possibly driving behavior. What happens between the visual input and the final action output is what needs to be understood. The visual world is firstly processed through the retina, a highly organized neural extension of the brain. The axons of retinal ganglion cells bundle together to form the optic nerve which contains fibers conveying features from the visual scene. Distinct representations of the visual world are detected in the retina, usually called features. Several feature detectors have been identified in distinct ganglion cell types, namely direction selective, object motion sensitive, looming detectors and Y versus X cells. Ganglion cell axons project to several brain regions and visual information is transmitted from the eye to the brain. How these different representations emerge, what computations are performed and which information is conveyed to the brain is a matter of investigation. When light reaches the retina a cascade of processing events happens that ultimately leads to vision. Light acts on photoreceptors, which in turn excite or inhibit bipolar cells resulting in ganglion cell activity changes. Since the publication "What the frog's eye tells the frog's brain" [2] the concept of feature detection in processing the information from the visual scene has been regarded as crucial in understanding how visual information is integrated. In humans retina axonal projections synapse in several regions of the brain, such as the lateral geniculate body, superior colliculi, pretectal nuclei, and the suprachiasmatic nucleus [3]. These projections convey information from distinct representations of the visual world that emerge through the interaction of different cell types present at the retina, namely photoreceptors, bipolar, horizontal, amacrine and ganglion cells. When light reaches the retina cones and rods transduce the information from the visual scene and this information is transmitted through different types of bipolar cells, namely ON and OFF bipolar cells which can act either as transient (high frequency) or sustained (low frequency) bipolar cells. The information contained in these parallel temporal channels is then conveyed to ganglion cells, the output cells of the retina. Another important player are amacrine cells that are involved in sharing information with ganglion cells leading to a correlated ganglion cell firing activity [4]. Moreover, and in order to be able to understand which computations are performed at the retina level it is important to have a definition for computation. According to Ian Horswill "The notions of input and output are the most basic concepts of computation: computation is for the moment the process of deriving the desired output from a given input(s)" [5]. Bearing this in mind how can we determine which are the computations performed in order to transform visual information carried by light into an electrophysiological signal? With the goal of answering to this question several animal models are being used including the frog, *drosophila melanogaster*, zebrafish, mice and primates.

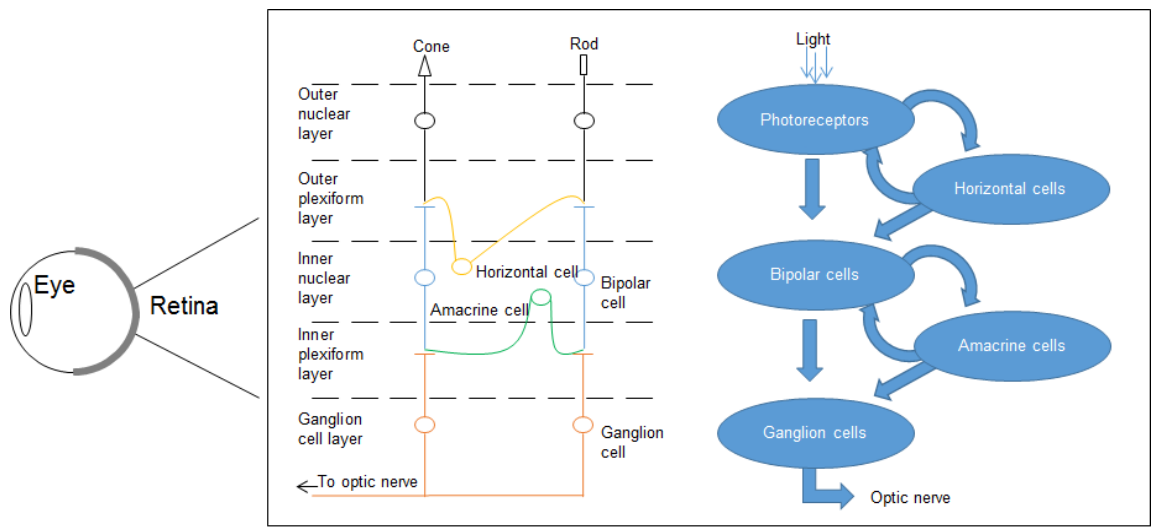

Figure 1 - Schematic representation of retinal layers and cell types.

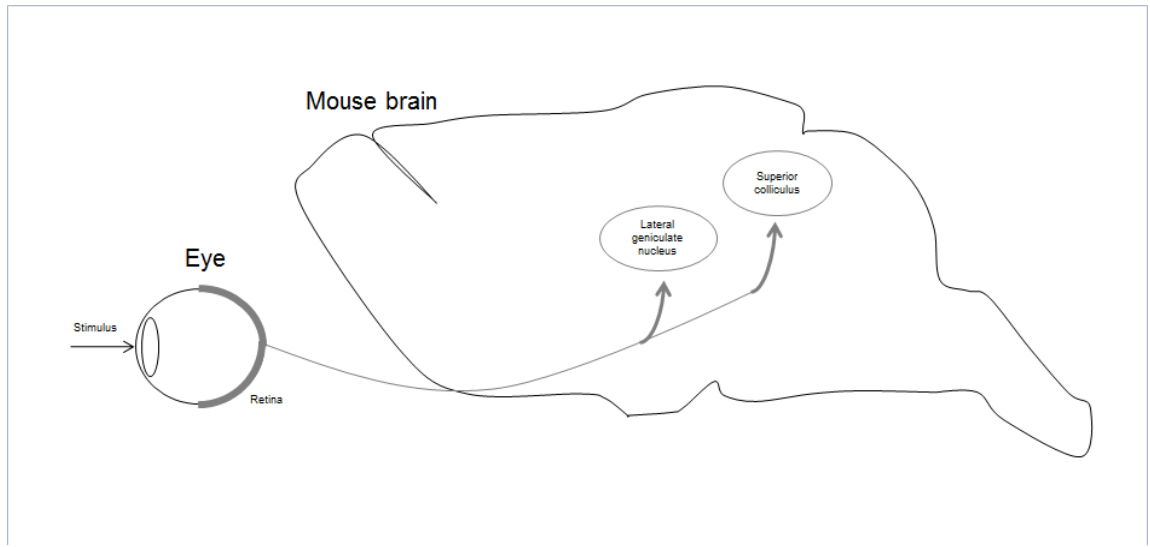

Figure 2 - Retinal projections to the lateral geniculate nucleus and superior colliculus.

The use of the frog as a model to study the retina can be traced back to the seminal paper of Lettvin and colleagues [2]. In this paper the frog is presented as having a simple visual system in which retinal projections target the superior colliculus, enabling to understand how visual features are transmitted to the brain. Since then several studies have been performed in different systems in order to try to understand how these features are encoded in different species. Moreover, which computations are performed in order to extract specific information such as the "bug perceiver" described by Lettvin and colleagues is still an area of intensive research. Similarly to frogs, zebrafish retinal axons project to the optic tectum [6]. Nikolau et al. studied the functional map of retinal projections to the tectum using a calcium fluorescent reporter specific for retinal ganglion cells and drifting gratings as visual stimuli. Different functional types of ganglion cells have been identified, namely direction selective and orientation selective [7]. Motion detection has also been studied for more than fifty years using *drosophila melanogaster* as a model [8]. Direction selective cells have been identified in flies [9]. Recently mice have started to be used as a visual model due to available genetic tools which enable to study neural circuits of vision. Mouse retina is similar to primate retina in the peripheral region which makes it an interesting model for low-acuity vision research [10]. The primate retina is able to perform various visual computations and visual processing is done through parallel channels which seems to start at the bipolar cell level. From the several types of neurons present at the retina it seems that only retinal ganglion cells (RGCs) and amacrine cells fire action potentials, an important issue in order to understand how visual information is processed and then integrated into other brain regions. This might indicate that the temporal processing present at the synaptic circuitry which communicates with ganglion cells somehow leads to the computations that might be modeled through a low-pass/band-pass filter because neurons at the retina, other than RGCs or amacrine cells, have graded potentials. Moreover, amacrine cells may play a role in synchronized activity in the retina since they have large synaptic outputs with RGCs [11].

## 2. Electrophysiology

Electroretinography and patch-clamp are two techniques that can be used to assess the electrical response of retinal neurons. While electroretinography expresses the concerted activity of neurons in the retina at a population level (figure 3), patch-clamp is used to determine the neural response at a single cell level (figure 4).

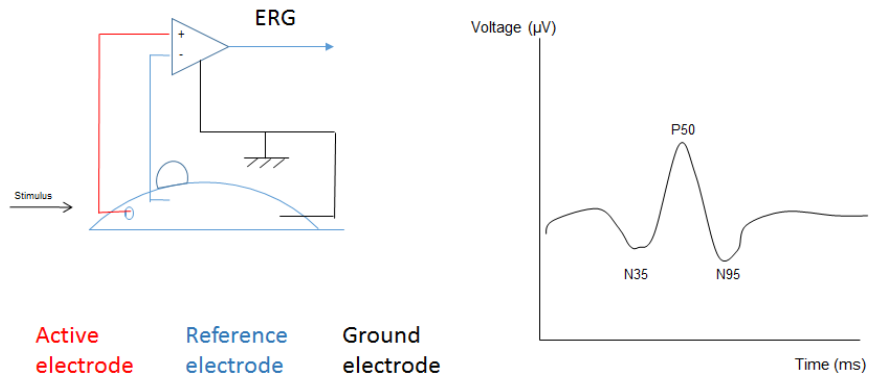

Figure 3 - Electroretinogram schematic setup representation.

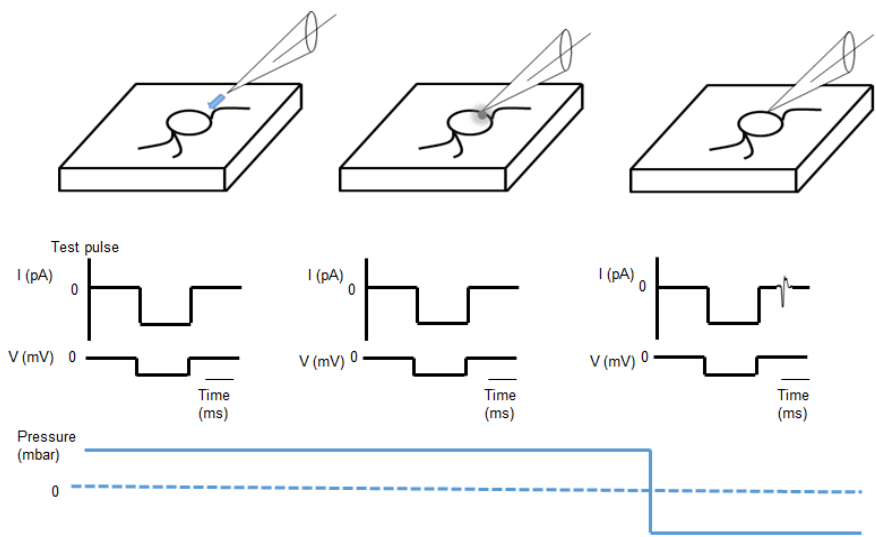

Figure 4 - Patch-clamp schematic setup representation of steps used for obtaining a recording. Top: whole-mount retina and patch pipette. Middle: oscilloscope display with a voltage test pulse. Bottom: pressure in the patch pipette.

## 3. Oscillators

A harmonic oscillator executes a simple harmonic movement in which for instance a given object moves around an equilibrium position and, in case there are not energy losses, the oscillation will remain. The oscillation of a body linked to a spring can be viewed as a harmonic oscillator while the gravity pendulum only behaves as a harmonic oscillator under specific conditions. In the following sections some examples of linear and nonlinear oscillators will be presented.

### 3.1. Linear oscillators

Let us start by considering the motion of a spring that is characterized by Hooke´s law. This law represents a simple harmonic motion oscillator.

#### 3.1.1. Spring

Hooke´s law

$$\overrightarrow{F_{elastic}} = -kx\overrightarrow{e_x}$$

Where F is the elastic force, k is the elastic constant and x is the displacement.

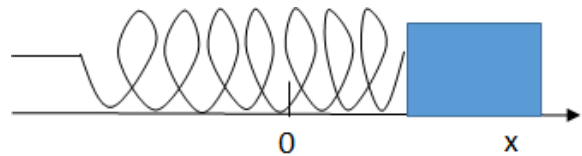

Figure 5 - Hooke´s law.

$$F = -kx = ma \Leftrightarrow m\frac{d^2x}{dt^2} = -kx$$

Where m is the body mass and a the acceleration.

$$\frac{d^2x}{dt^2} = -\frac{k}{m}x = -\omega^2 x$$

Where w is the angular frequency.

The solution is $x(t) = A\sin(\omega t + \varphi_0)$, a harmonic oscillator, where A is the oscillation amplitude.

$\varphi = \omega t + \varphi_0$ is the phase, where $\varphi_0$ is the initial phase.

$\omega = \frac{2\pi}{T} = 2\pi f$ is the angular frequency, T is the period and f the frequency.

$$v = \frac{dx}{dt} \Leftrightarrow v = A\cos(\omega t + \varphi_0)$$

$$a = \frac{dv}{dt} \Leftrightarrow a = -A\omega^2 \sin(\omega t + \varphi_0) = -\omega^2 x$$

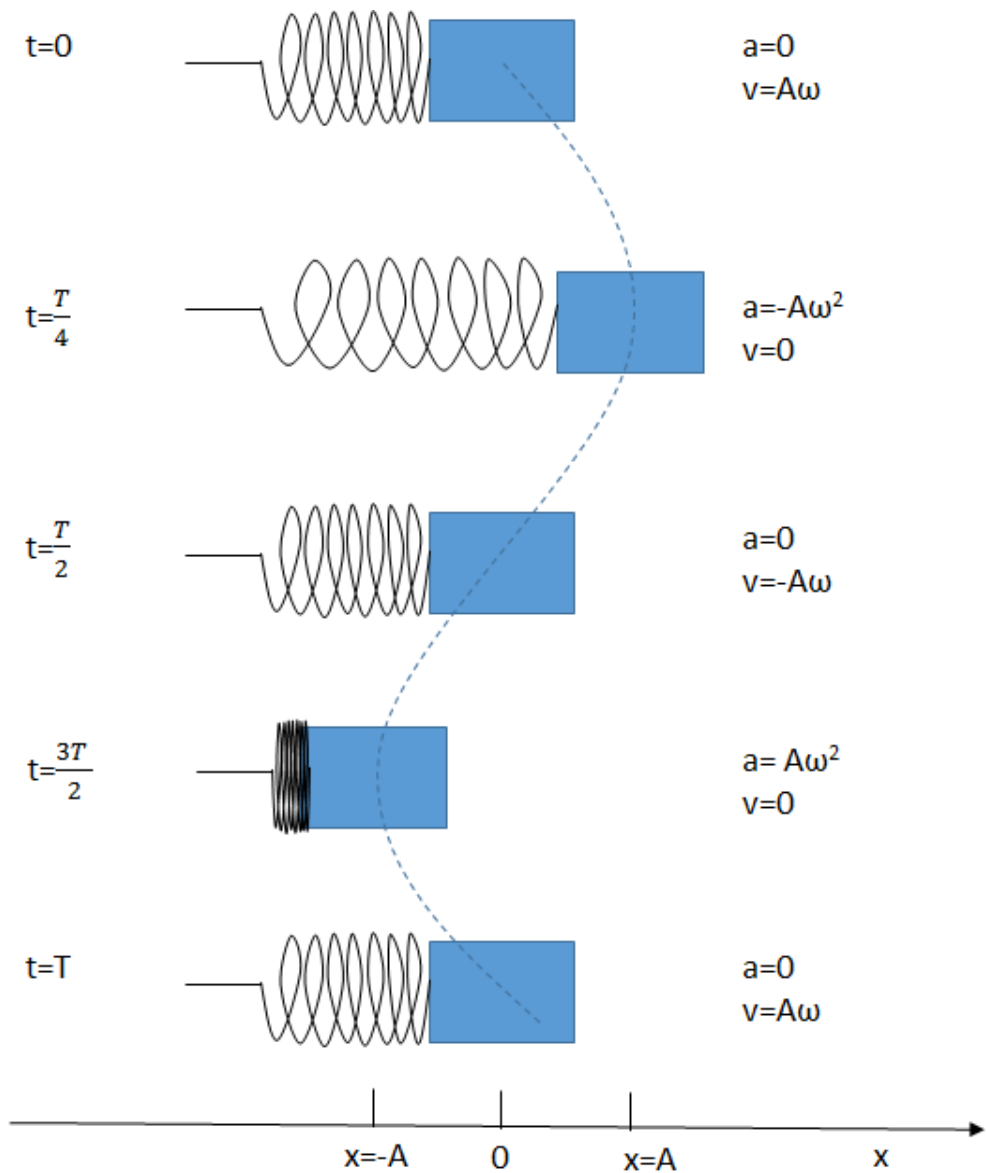

Figure 6 - Harmonic oscillation.

3.1.2. Pendulum

The gravity pendulum is another example of a model system that can behave as a harmonic oscillator.

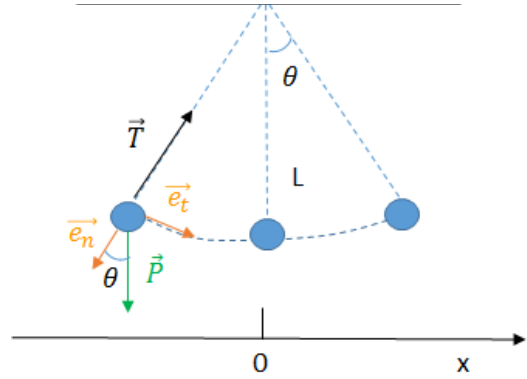

Figure 7 - Force diagram of a gravity pendulum.

$$\begin{bmatrix} \Sigma F_t = ma_t \\ \Sigma F_n = ma_n \end{bmatrix} \Leftrightarrow \begin{bmatrix} -mg\sin\theta = ma_t \\ T - mg\cos\theta = ma_n \end{bmatrix} \Leftrightarrow \begin{bmatrix} -g\sin\theta = a_t \\ T - mg\cos\theta = ma_n \end{bmatrix}$$

Where L is the length of the wire with tension T; P is the pendulum body weight; g is the gravity constant.

For $\theta \leq 10°$ the tangential component of acceleration is around $a_t = -g\sin\theta = -g\theta$ because $\sin\theta \cong \theta$.

For small oscillations the gravity pendulum can be considered a harmonic oscillator $a_t = -g\theta \Rightarrow a_t = -\frac{g}{L}x$.

Since $a = -w^2 x$, $\omega^2 = \frac{g}{L}$ and $f = \frac{1}{T} = \frac{1}{2\pi}\sqrt{\frac{g}{L}} \Leftrightarrow T = 2\pi\sqrt{\frac{L}{g}}$.

3.2. Nonlinear oscillators

Nonlinearity is commonly observed in Biology. In nonlinear systems a sinusoidal input can give rise to an output response containing the fundamental frequency and higher harmonics. The Hodgkin-Huxley model of action potentials is an example of such a system [12]. The Hodgkin-Huxley model provides a mathematical model for the action potential generation in which sodium and potassium play a fundamental role [13].

3.2.1. The Hodgkin-Huxley model

After a stimulus sodium influx and potassium outflux can lead to an action potential. The electrical circuit proposed in 1952 by Hodgkin and Huxley is depicted in figure 8.

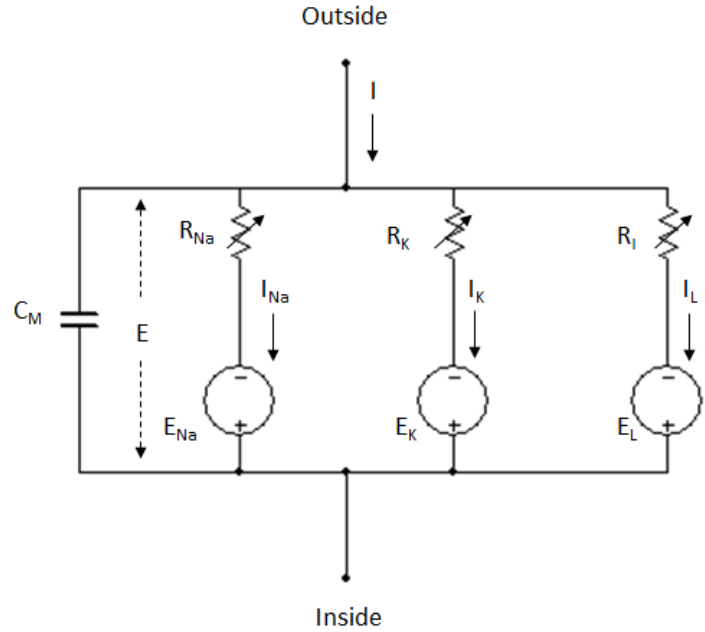

Figure 8 - Hodgkin-Huxley electrical circuit of nerve membrane.

The following equation relates the membrane current I with membrane potential E:

$$I = C\frac{dV}{dt} + g_K(E - E_K) + g_{Na}(E - E_{Na}) + g_L(E - E_L)$$

where $I_{Na}$ and $I_K$ represent sodium and potassium current while $I_L$ represents the leakage current. Conductances for sodium, potassium and leakage are represented by $g_{Na} = \frac{1}{R_{Na}}$, $g_K = \frac{1}{R_K}$, $g_L = \frac{1}{R_L}$. The membrane capacitance is represented by $C_M$ and sodium, potassium and leakage potentials are represented by $E_{Na}$, $E_k$, $E_L$. The equations that summarize the Hodgkin-Huxley model are the following:

$$\frac{dV_m}{dt} = -\frac{1}{C_m}(g_{Na}m^3h(V_m - E_{Na}) + g_K n^4(V_m - E_K) + g_L(V_m - E_L) - I_{stim})$$

$$\frac{dm}{dt} = \alpha_m(1-m) - \beta_m m$$

$$\frac{dh}{dt} = \alpha_h(1-h) - \beta_h h$$

$$\frac{dn}{dt} = \alpha_n(1-n) - \beta_n n$$

$$\alpha_m = \frac{0.1(-35 - V_m)}{\exp\left(\frac{-35 - V_m}{10}\right) - 1}$$

$$\beta_m = 4\exp\left(\frac{-60 - V_m}{18}\right)$$

$$\alpha_h = 0.07\exp\left(\frac{-60 - V_m}{20}\right)$$

$$\beta_h = \frac{1}{exp\left(\frac{-30 - V_m}{10}\right) + 1}$$

$$\alpha_n = \frac{0.01(-50 - V_m)}{exp\left(\frac{-50 - V_m}{10}\right) - 1}$$

$$\beta_n = 0.125exp\left(\frac{-60 - V_m}{80}\right)$$

where $V_m$ is the membrane potential; $C_m$ is the membrane capacitance; $g_{Na}$, $g_K$, $g_L$ are the sodium, potassium and leakage conductances $m$, $h$, $n$ are the gating variables; $E_{Na}$, $E_K$, $E_L$ are Nernst potentials for sodium, potassium and leakage, respectively; $\alpha$, $\beta$ are the rate coefficients. With these equations it is possible to implement a code in Matlab in order to see the membrane potential response to a series of current stimulations (figure 9) [14, 12].

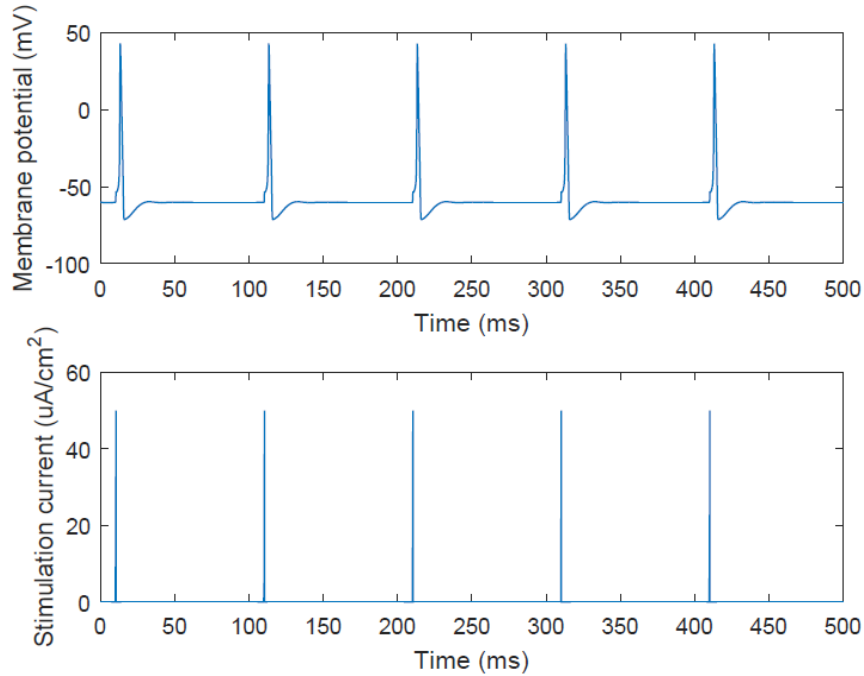

Figure 9 - Membrane potential response to a series of current inputs implemented according to the Hodgkin-Huxley model.

## 4. Fourier analysis

Fourier transform can be used to analyze the response of a circuit to sinusoidal inputs. Fast Fourier transform algorithm enables the practical implementation of Fourier transform in order to analyze signals and systems. This mathematical operation was discovered by Joseph Fourier and can be used to describe periodic variations in time. Fourier analysis can be applied when there is a sinusoidal phenomenon in time in which we want to try to understand which their sinusoidal components are. The Fourier series of a continuous-time periodic signal is represented by the following pair of equations:

$$x(t) = \sum_{-\infty}^{+\infty} a_k e^{jk\omega_0 t} = \sum_{-\infty}^{+\infty} a_k e^{jk(2\pi/T)t}$$

$$a_k = \frac{1}{T}\int_T x(t)e^{-jkw_0 t}dt = \frac{1}{T}\int_T x(t)e^{-jk(2\pi/T)t}dt$$

These expressions represent the synthesis and the analysis equations. In these equations $w_0$ represents the fundamental frequency, $T$ represents the fundamental period.

The Fourier series of discrete-time periodic signals is represented by the discrete-time Fourier series pair:

$$x[n] = \sum_{K=\langle N \rangle} a_k e^{jk\omega_0 n} = \sum_{K=\langle N \rangle} a_k e^{jk(2\pi/N)n}$$

$$a_k = \frac{1}{N}\sum_{n=\langle N \rangle} x[n]\, e^{-jk\omega_0 n} = \frac{1}{N}\sum_{n=\langle N \rangle} x[n]\, e^{-jk(2\pi/N)n}$$

where $N$ is the fundamental period, $\omega_0 = 2\pi/N$ is the fundamental frequency and these expressions representing the synthesis and the analysis equations.

The Fourier transform can be applied in aperiodic signals. This signal can be understood as a periodic signal with infinite period. The inverse Fourier transform and The Fourier transform take the following form:

$$x(t) = \frac{1}{2\pi}\int_{-\infty}^{+\infty} X(j\omega)\, e^{j\omega t}d\omega$$

$$X(j\omega) = \int_{-\infty}^{+\infty} x(t)e^{-jwt}dt$$

The discrete time Fourier transform has the following description [15]:

$$x[n] = \frac{1}{2\pi}\int_{2\pi} X(e^{jw})e^{jwn}dw$$

$$X(e^{j\omega}) = \sum_{n=-\infty}^{n=+\infty} x[n]e^{-j\omega n}$$

## 5. Retina model

Uniform field flashes and grating patterns can be used to investigate the origin of ON versus OFF responses in ganglion cells [16]. When a stimulus with a specific frequency is applied to the retina the response output is modulated with a specific frequency output. With this approach it is possible to determine the characteristic response of the system we are studying, the retina. Therefore, by using a sine-wave modulated light or flicker input the temporal response signature of the retina can be assessed. In a nonlinear system the response to several sine-wave light inputs contains components at the input frequencies, at the difference, the sum and the multiples of the input frequencies [17].

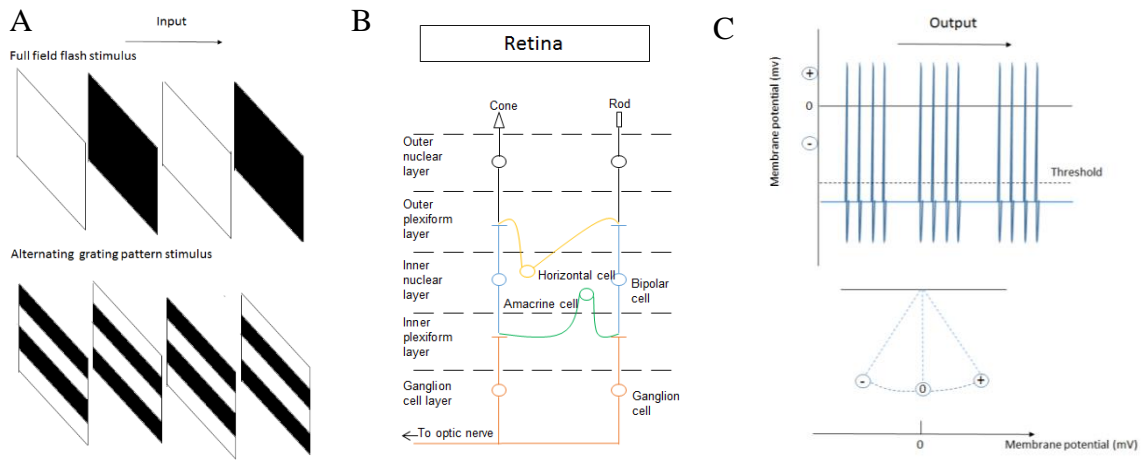

Figure 10 - A) Stimuli for flicker ERG and pattern ERG; B) Retina block diagram; C) Action potential in response to square wave visual stimulus which leads to an oscillatory response.

From figure 10 we can think that the retina can resemble an oscillator. Nevertheless, whether this oscillator is linear or nonlinear is a question to take into consideration. Bearing in mind the Hodgkin-Huxley nonlinear model we can guess that it is possibly a nonlinear system. In the following sections we will go through linear versus nonlinear systems with reference to original work showing that the retina behaves as a nonlinear system.

## 5.1. The retina as a nonlinear system

Graded potentials imply that high frequency components of the visual scene might not be represented at the level of photoreceptor, horizontal and bipolar cells. Moreover, due to the spikes fired by amacrine and ganglion cells there is an all or nothing mechanism in which action potentials translate the information from the visual world into a new code which is then transmitted to other brain areas through the optic nerve. There a nonlinear mechanism might be present due to rectification and summation implemented by amacrine and ganglion cells. The system shown in figure 11 receives an input u that is converted through a linear filter resulting in x. The output of the linear filter is the input for the nonlinear system with the output y which can contain a fundamental frequency and its harmonics. This output might be represented by the Fourier series:

$$x(t) = X_0 + \sum_{n=1}^{\infty}(a_n sinwt + b_n coswt)$$

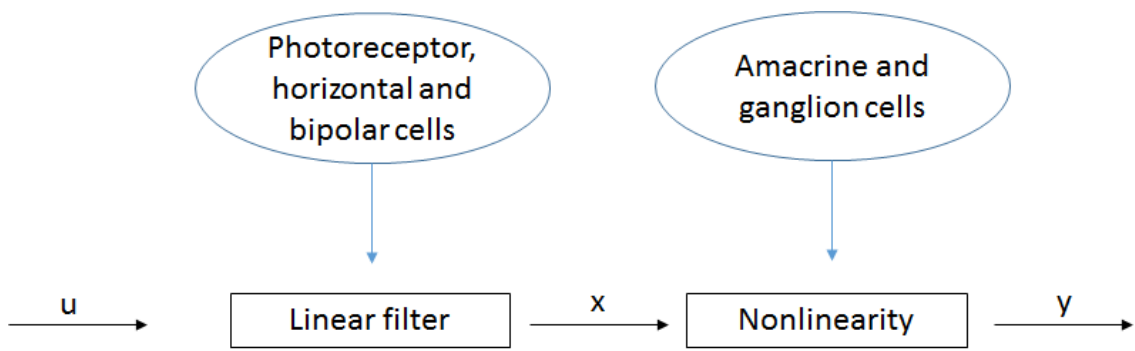

Figure 11 - Retina block diagram

The stimulus that arrives at the retina is integrated with a spatiotemporal linear filter, then a threshold nonlinear mechanism and noise are present which leads to the output spike trains [11].

## 5.1.1. Linear filter

A filter is able to change the amplitude and frequency components of a given signal. A low-pass or high-pass filter can be implemented through a RC circuit as shown in figure 12.

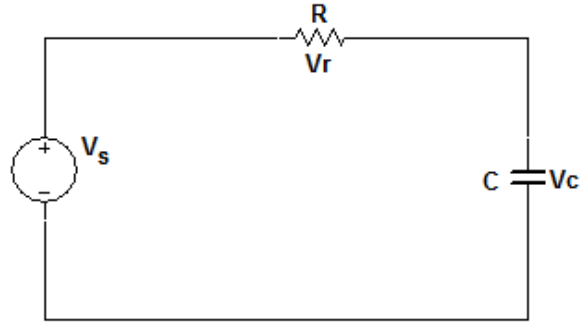

Figure 12 - RC filter circuit.

The capacitor voltage can be determined according to the following expression:

$$RC\frac{dv_c}{dt} + v_c(t) = v_s(t)$$

Assuming an input voltage of $v_s(t) = e^{jwt}$ the output voltage of the system $H(jw)$ is $v_c(t) = H(jw)e^{jwt}$. Then the previous equation becomes

$$RC\frac{d}{dt}[H(j\omega)e^{j\omega t}] + H(j\omega)e^{j\omega t} \Leftrightarrow RCj\omega H(j\omega)e^{j\omega t} + H(j\omega)e^{j\omega t} = e^{j\omega t}$$

which can be expressed as

$$H(j\omega) = \frac{1}{1 + RCj\omega}$$

This RC circuit with $v_c(t)$ output is a low-pass filter. Instead of looking to the output of the capacitor we can look to the output voltage of the resistor, which depicts a high-pass filter.

$$RC\frac{dv_r(t)}{dt} + v_r(t) = RC\frac{dv_s(t)}{dt}$$

With $v_r(t) = G(jw)e^{jwt}$ we get [15]

$$G(j\omega) = \frac{j\omega RC}{1 + j\omega RC}$$

A Reichardt detector can be implemented by combining a low-pass filter with a high-pass filter, as described by Weber and colleagues [18].

### 5.1.2. Rectification and Summation

Electrical coupling between neurons can lead to preferential current flow in one direction, a process defined as rectification [19]. Rectifying electrical synapses transmit depolarizing signals in one direction and hyperpolarizing signals in the opposite direction. In other words, rectifying synapses show preferential flow of positive current in one direction. In case there is more than one synchronized presynaptic input to a postsynaptic neuron the resulting postsynaptic currents sum, which is termed as summation [20]. In the computation of direction selectivity summation plays an important role. This has been observed in *drosophila melanogaster* lobula plate tangential cells that compute direction selectivity through summation of inputs with distinct direction preference [9].

5.1.3. Feature detectors

In motion detection a temporal filter and nonlinear mechanism are present enabling to determine the direction of motion. This computation starts at the photoreceptor level, where the brightness of two adjacent points is compared, goes to the bipolar cells, where the input information is separated into several parallel channels, and continues to the ganglion cell layer, where information from several bipolar cells is integrated enabling to compute a specific feature of the visual scene. Amacrine and horizontal cells also play an important modulatory role. Retinal ganglion cells encode the information from the visual scene in the spatiotemporal pattern of action potentials. Furthermore, several distinct computations can result from the interaction of retinal cell types namely direction selective, object motion sensitive, looming detectors and Y versus X cells. X ganglion cells have a linear receptive field with sustained response while Y cells have a nonlinear receptive field with transient changes in firing [4, 11]. The basic building blocks seem to include temporal filtering and nonlinear rectification mechanisms.

5.1.4. Retina equivalent circuit

An electrode patched in a cell-attached configuration can be used in order to accurately report the membrane potential of a ganglion cell. Cell-attached current-clamp recordings can be used to determine resting and synaptic potentials. The voltage $V_0$ measured with zero current input is the following (figure 13):

$$V_0 = \frac{E_m \times R_{seal}}{R_{seal} + R_{patch+cell}}$$

where $E_m$ is the resting membrane potential; $R_{seal}$, $R_{patch}$, $R_{cell}$, $R_{electrode}$ are the resistances from the seal, patch, cell and electrode. With $ratio = R_{seal}/R_{patch+cell}$

$$\frac{V_0}{E_m} = \frac{ratio}{(1 + ratio)}$$

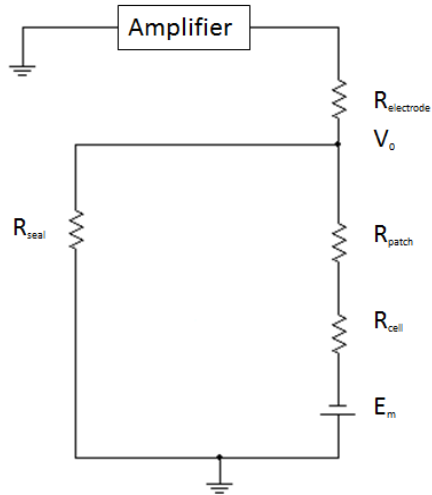

Figure 13 - Voltage measured by an amplifier in cell-attached current clamp mode with zero current input.

When the cell membrane is changing the membrane patch can be represented by a resistor and a capacitor in parallel (figure 14). In cell-attached current clamp mode the capacitor behaves as a low-pass filter. With an infinite seal resistance we can assume that the pipette voltage ($V_p$) is equal to the membrane voltage ($V_m$). Therefore,

$$\Delta V_p = \Delta V_m (1 - e^{-t/\tau})$$

where $\tau = R_{patch} \times C_{patch}$ [21]. As we will see this simple cell membrane circuit can be used to model retinal feature detection, assuming that potentials at the bipolar-RGCs synapses can accurately depict the information that is going to be transmitted to other brain regions through RGCs axons.

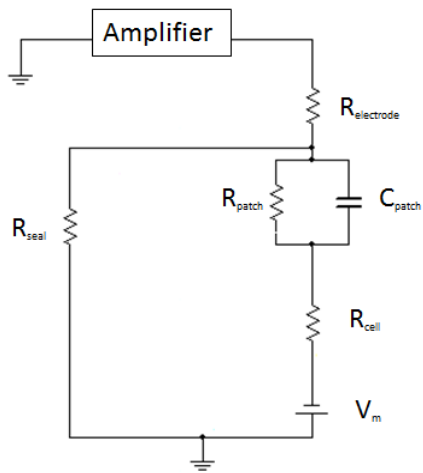

Figure 14 - Cell membrane circuit with changing potential.

### 5.1.5. Retina feature detector model

Munch et al. have described a neural circuit for implementing motion detection. A computational model of a linear-nonlinear system for approaching objects was proposed [22]. The response of PV-5 ganglion cells was modeled according to the following equation

$$-C_m \frac{dV_m}{dt} = \frac{(V_m - V_{rest})}{R_m} + g_{ex}(t)(V_m - V_{ex}) + g_{inh}(t)(V_m - V_{inh})$$

where $C_m$ is the cell capacitance; $R_m$ is the cell resistance; $V_{exc}$ is the reversal excitatory input potential; $V_{inh}$ is the reversal inhibitory input potential; $V_{rest}$ is the resting membrane potential; $g_{ex}$ and $g_{inh}$ are the excitatory and inhibitory conductances of ganglion cells. Looking to the previous equation we can easily see that this equation corresponds to a capacitor in parallel with a resistor through which current $i$ flows, which can be described as follows

$$i_c = i_r + i_{ex} + i_{inh}$$

### 5.1.6. Retina model simulation

Besides representing ganglion cells with passive mechanisms such as RC parallel circuits it is possible to implement a nonlinear Hodgkin-Huxley model with passive dendrites. Using NEURON windows version 7.4 [23] a model of amphibian ganglion cells can be simulated [24]. The aim of the simulation that is going to be presented is to understand the voltage response to a current point source located at the soma and to gain some insights into the possible synchronous output in invertebrate neurons. After soma depolarization the action potential at the axon output was analyzed. The ganglion cell was simulated according to the parameters used by Greenberg and colleagues [24]. A Hodgkin-Huxley model was set for the soma and axon, while the dendrites were simulated by a passive mechanism. The simulation was performed with temperature of 22 ºC, resting membrane potential of -70 mV, time step of 25 $\mu s$. The passive dendrites were simulated with membrane capacitance of $1\ \mu F/cm^2$, membrane resistance of 50000 $\Omega cm^2$, cytoplasmic resistance of 110 $\Omega cm$. The active nonlinear Hodgkin-Huxley soma and axon were simulated with sodium, potassium and leak conductances $g_{Na} = 0.120\ S/cm^2$, $E_{Na} = 50\ mV$, $g_K = 0.036\ S/cm^2$, $E_K = -77\ mV$, $g_L = 0.0003\ S/cm^2$, $E_L = -54.3\ mV$. The soma diameter was $d = 24\ \mu m$; axon length was $l = 1000\ \mu m$ with variable diameter between $d = 0.6:1.2\ \mu m$; dendrite length was $l = 200\ \mu m$ with variable diameter between $d = 10:3\ \mu m$. A point source was placed over the soma with regular stimulus injections with delay of $del = 100\ ms$, duration of $dur = 100\ ms$, amplitude of $amp = 50\ nA$ and frequency of $freq = 5\ Hz$. The axon potential at the axon output of one cell is depicted in figure 15. A new series of stimulus injections was provided with delay of $del = 100, 500\ ms$, duration of $dur = 5\ ms$, amplitude of $amp = 50\ nA$ and frequency of $freq = 2, 5\ Hz$. The result is present in figure 16. With this simple example we can see that two different input frequencies can originate an output with two different output frequencies.

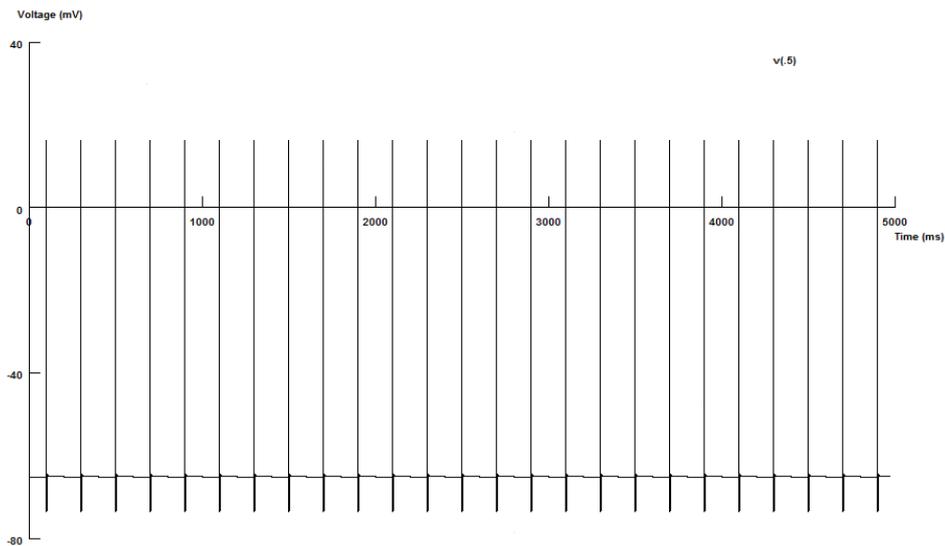

Figure 15 - Action potential at ganglion cell axon output.

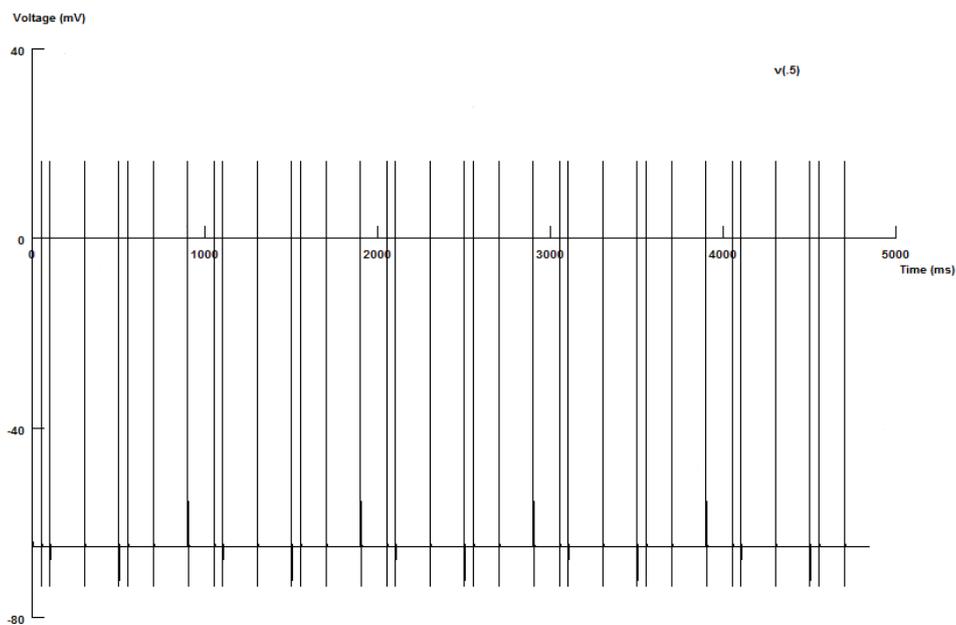

Figure 16 - Action potential at ganglion cell output with two input signals at distinct frequencies.

In linear systems a sinusoidal perturbation results in a sinusoidal output with the same frequency (figure 17). Let us consider different sinusoidal input frequencies. In a linear system the response to each frequency is present in the output and, if both inputs are present, both output frequencies are also present. This is the example presented in figure 17 in which two input sine wave frequencies of 1 and 3 Hz give rise to an output with the same frequencies. Let us now consider a nonlinear system. In nonlinear systems a sinusoidal input can give rise to a response with the fundamental frequency and higher harmonics (figure 18). This is the case of the retina, as demonstrated by Chang and colleagues [25]. In this paper the authors show that the response of the retina to flicker electroretinogram is nonlinear. Therefore, the response to a single frequency input can have higher harmonics and if several input frequencies are present several output harmonics are observed. The example of figure 18 represents a raster plot of two input sine wave frequencies that are corrupted with random noise and to which is assigned the value of 0 when there is any spike or the value of 1 when there is a spike. Besides the input frequencies of 1 and 3 Hz, the Fourier transform shows peaks at 2, 4 and 6 Hz. This means that by adding random noise and threshold to the input signal a nonlinear behavior can emerge.

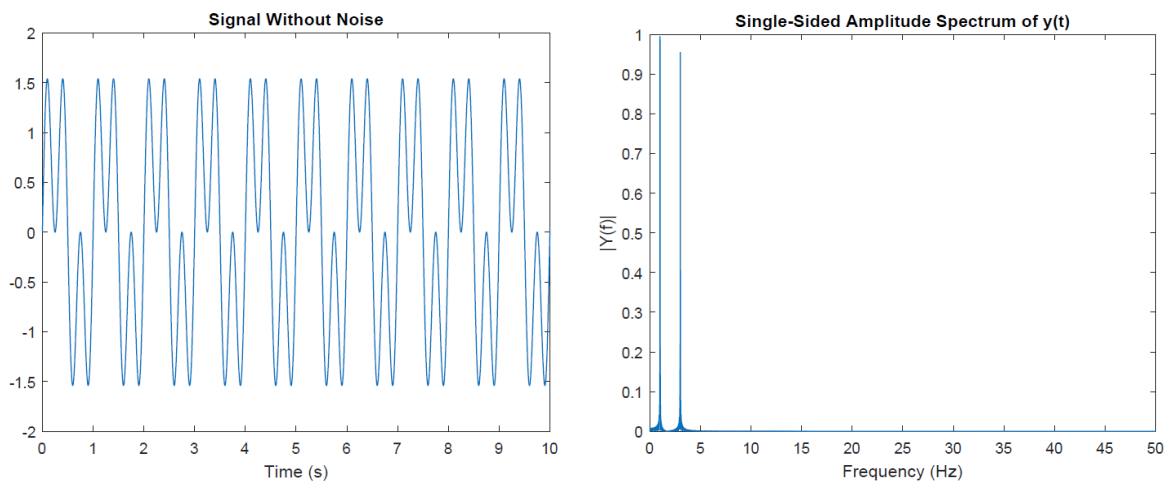

Figure 17 - Linear system representation.

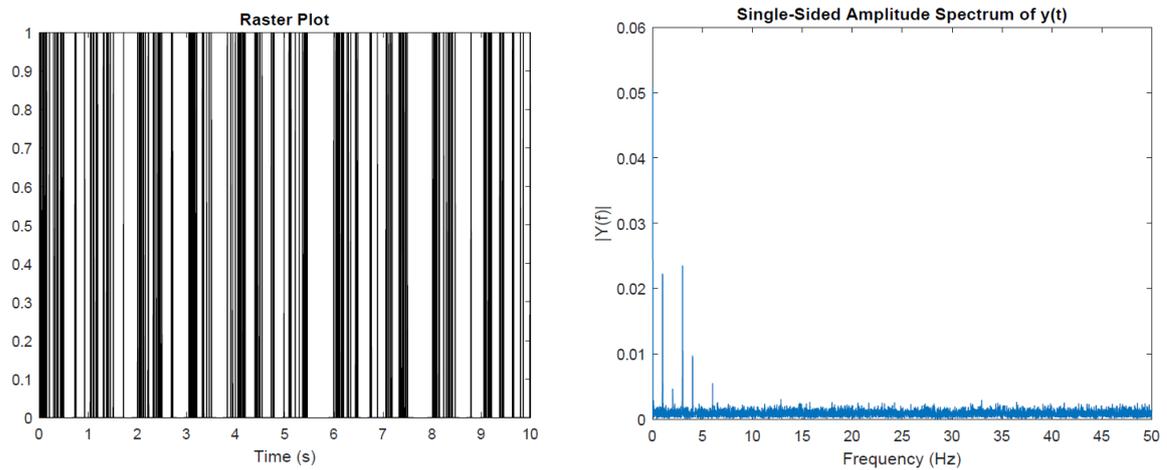
Figure 18 - Nonlinear system representation.

## 6. Open questions

Saint Augustine once said "What then is time? If no one asks me, I know. If I wish to explain to him who asks, I know not". This question raises some philosophical thoughts about what is time and how can we quantify it. The notion that time is associated with movement is old and the use of a gravity pendulum in clocks is an example of such an observation. Albert Einstein went deeper into this question and showed in his general relativity theory that space and time can be viewed as the sides of the same coin. Time might be regarded as a change in space, thus one cannot exist without the other. This leads to the question on how space and time are integrated in the visual system into higher cortical areas. The frequency doubling technique shows us that these apparently distinct aspects of reality, space and time, can be easily mixed. Future investigations will clarify on how retinal computations are transmitted to higher visual areas and which computations are performed in order to reach what we call the visual world.

## 7. Acknowledgements

SC was supported by doctoral fellowship SFRH/BD/52045/2012 from Fundação para a Ciência e a Tecnologia (FCT).

# 9. Appendix

## 9.1. MatLab code for the Hodgkin-Huxley model

```matlab
%%HHcaller.m
clear all
clc

options=odeset('RelTol',1e-10,'AbsTol' ,[1e-10 1e-10 1e-10 1e-10],'MaxStep' ,0.05); % precision settings for integrator

global input_start input_stop input_magnitude input_start1 input_stop1
input_magnitude1 input_start2 input_stop2 input_magnitude2
input_start3 input_stop3 input_magnitude3 input_start4 input_stop4
input_magnitude4

input_start=10; % ms
input_stop=10.3; % ms
input_magnitude=50; % uA/cm^2
input_start1=110; % ms
input_stop1=110.3; % ms
input_magnitude1=50; % uA/cm^2
input_start2=210; % ms
input_stop2=210.3; % ms
input_magnitude2=50; % uA/cm^2
input_start3=310; % ms
input_stop3=310.3; % ms
input_magnitude3=50; % uA/cm^2
input_start4=410; % ms
input_stop4=410.3; % ms
input_magnitude4=50; % uA/cm^2
% Define constants
V=-60; % Initial membrane voltage, mV

% Alpha and beta parameters
am=0.1*(V+35)/(1-exp(-(V+35)/10));%Alpha for Variable m
bm=4.0*exp(-0.0556*(V+60)); %Beta for variable m
an=0.01*(V+50)/(1-exp(-(V+50)/10));%Alpha for variable n
bn=0.125*exp(-(V+60)/80); %Beta for variable n
ah=0.07*exp(-0.05*(V+60));%Alpha value for variable h
bh=1/(1+exp(-(0.1)*(V+30)));%beta value for variable h
% Initialize state variables
m=am/(am+bm); % Initial m-value
n=an/(an+bn); % Initial n-value
h=ah/(ah+bh); % Initial h-value
%% Vall Runge-Kutta 4 ODE

[time v]=ode45(@HHmodel, [0 500],[V n m h] , options); % integrates
model from initial conditions vector

%% Calculate  Is-external current applied, microamps/cm^s
for time_step=1:length(time)
    if input_start<=time(time_step) && time(time_step)<input_stop
        I_s(time_step)=(time(time_step)-input_start).*input_magnitude/(input_stop-input_start);
    elseif input_start1<=time(time_step) && time(time_step)<input_stop1
```

```matlab
            I_s(time_step)=(time(time_step)-
input_start1).*input_magnitude1/(input_stop1-input_start1);
        elseif input_start2<=time(time_step) && 
time(time_step)<input_stop2
            I_s(time_step)=(time(time_step)-
input_start2).*input_magnitude2/(input_stop2-input_start2);
        elseif input_start3<=time(time_step) && 
time(time_step)<input_stop3
            I_s(time_step)=(time(time_step)-
input_start3).*input_magnitude3/(input_stop3-input_start3);
          elseif input_start4<=time(time_step) && 
time(time_step)<input_stop4
            I_s(time_step)=(time(time_step)-
input_start4).*input_magnitude4/(input_stop4-input_start4);
    else
        I_s(time_step)=0;
    end
end

%% Calculate values to find currents
R=8.31447; % J K^-1 mol ^-1
F=9.6485309*10^4; % C mol^-1
Temp=279.45; % Kelvin
V_Na=((R*Temp)/F) * log(490/50)*10^3;%Na reversal potential, mV
V_K=((R*Temp)/F) * log(20/400)*10^3; %K reversal potential, mV
V_L=-50;%Leakage reversal potential, mV
G_Na_max=120; % % Na conductance, mS/cm^2
G_K_max=36; %K conductance, mS/cm^2
G_Na=G_Na_max.*v(:,3).^3.*v(:,4);
G_K=G_K_max.*v(:,2).^4;
G_L=.3; % %Leakage conductance, mS/cm^2
I_Na=(v(:,1)-V_Na).*G_Na;% uA/cm^2
I_K=(v(:,1)-V_K).*G_K;% uA/cm^2
I_L=(v(:,1)-V_L).*G_L;% uA/cm^2

%% Figure plots
figure(1)
subplot(2,1,1)
plot(time, v(:,1))
ylabel('Membrane potential (mV)')
xlabel('Time (ms)')
subplot(2,1,2)
plot(time, I_s)
ylabel('Stimulation current (uA/cm^2)')
xlabel('Time (ms)')

%%HHmodel.m
function [dv] = HHmodel(t, v)

global input_start input_stop input_magnitude input_start1 input_stop1 
input_magnitude1 input_start2 input_stop2 input_magnitude2 
input_start3 input_stop3 input_magnitude3 input_start4 input_stop4 
input_magnitude4

%% Set up the constants
R=8.31447; % J K^-1 mol ^-1
F=9.6485309*10^4; % C mol^-1
Temp=279.45; % Kelvin

% Sax conductances
```

```matlab
G_Na_max=120; % mS/cm^2
G_K_max=36; % mS/cm^2

% Nernst potentials in mV
V_Na=((R*Temp)/F)*log(490/50)*10^3;
V_K=((R*Temp)/F)*log(20/400)*10^3;
V_L=-50;

% Membrane capacitance
C_mem=1;% Membrane capacitance, uF/cm^2

% V rest
Vr=-60; % mV

%% Time constants
a_m=.1*(25-(v(1)-Vr))/(exp((25-(v(1)-Vr))/10)-1);
a_n=.01*(10-(v(1)-Vr))/(exp((10-(v(1)-Vr))/10)-1);
B_m=4/exp((v(1)-Vr)/18);
B_n=.125/exp((v(1)-Vr)/80);
a_h=.07/exp((v(1)-Vr)/20);
B_h=1/(exp((30-(v(1)-Vr))/10)+1);

%% Conductances
G_Na=G_Na_max.*v(3).^3.*v(4);
G_K=G_K_max.*v(2).^4;
G_L=.3;

%% Currents
I_Na=(v(1)-V_Na).*G_Na;
I_K=(v(1)-V_K).*G_K;
I_L=(v(1)-V_L).*G_L;

%% Input ramp
if input_start<=t && t<input_stop
    Is=(t-input_start).*input_magnitude/(input_stop-input_start);
elseif input_start1<=t && t<input_stop1
    Is=(t-input_start1).*input_magnitude1/(input_stop1-input_start1);
elseif input_start2<=t && t<input_stop2
    Is=(t-input_start2).*input_magnitude2/(input_stop2-input_start2);
elseif input_start3<=t && t<input_stop3
    Is=(t-input_start3).*input_magnitude3/(input_stop3-input_start3);
    elseif input_start4<=t && t<input_stop4
    Is=(t-input_start4).*input_magnitude4/(input_stop4-input_start4);
else
    Is=0;
end

%% First order ordinary differential equations
dv=zeros(4,1);
dv(1)=-1/C_mem*(I_Na+I_K+I_L-Is); % membrane voltage
dv(2)=a_n*(1-v(2))-B_n*v(2); % n
dv(3)=a_m*(1-v(3))-B_m*v(3); % m
dv(4)=a_h*(1-v(4))-B_h*v(4); % h
```

## 9.2. Retina model simulation

```
create soma, axon, dendrite[3]
connect axon(0), soma(0)
for i=0,2 { connect dendrite[i](0), soma(1)}
access axon
soma nseg = 1
soma diam=24
soma L=24
soma Ra=110
soma insert hh
objectvar stim
soma stim=new IClamp(0.5)
stim.del=100
stim.dur=5
stim.amp=50
dt=0.025
tstop=5000
objectvar stim1
soma stim1=new IClamp(0.5)
stim1.del=300
stim1.dur=5
stim1.amp=50
objectvar stim2
soma stim2=new IClamp(0.5)
stim2.del=500
stim2.dur=5
stim2.amp=50
objectvar stim3
soma stim3=new IClamp(0.5)
stim3.del=700
stim3.dur=5
stim3.amp=50
objectvar stim4
soma stim4=new IClamp(0.5)
stim4.del=900
stim4.dur=5
stim4.amp=50
axon nseg = 1000
axon L = 1000
axon Ra=110
axon diam(0:1)=0.6:1.2
axon insert hh
for i=0,2 dendrite[i] {nseg = 200 L = 200 Ra=110 diam(0:1) = 10:3 insert pas
e_pas = -70 g_pas = 0.00002}
```

## 9.3. MatLab code for linear versus nonlinear system

```matlab
%----- Initiate -----%
clear all; close all;
time = 100; % seconds
samplingrate = 1000; % Hz
X = 0:1/samplingrate:time-1/samplingrate;
%----- Make Spike Train -----%
st = randn(10, time*samplingrate)-2.2;
signal = .5*sin(2*pi*1*X)+.5*sin(2*pi*3*X);
st(1,:) = st(1,:) + signal;%Signal Corrupted with Zero-Mean Random Noise
spikes = find(st > 0);
nospikes = find(st<=0);
st(nospikes) = 0;
st(spikes) = 1;
SpikeTrain = st;
%----- Plot Raster Plot -----%
figure(1)

plot(X, st(1,:),'k'); hold on
title ('Raster Plot')
xlim([0 10])
xlabel ('Time (s)')

%Fourier transform example
% Plot sine wave
X = 0:1/samplingrate:time-1/samplingrate;
y=sin(2*pi*1*X)+sin(2*pi*3*X);
figure(2)
plot(X,y)
xlim([0,10])
title('Signal Without Noise')
xlabel('Time (s)')
L=length(y)
NFFT = 2^nextpow2(L); % Next power of 2 from length of y
Y = fft(y,NFFT)/L;
f = samplingrate/2*linspace(0,1,NFFT/2+1);

% Plot single-sided amplitude spectrum.
figure(3)
plot(f,2*abs(Y(1:NFFT/2+1)))
title('Single-Sided Amplitude Spectrum of y(t)')
xlabel('Frequency (Hz)')
ylabel('|Y(f)|')
xlim([0 50])

%Fourier transform of sine waves corrupted with random noise
L1=length(y)
NFFT = 2^nextpow2(L1); % Next power of 2 from length of y
Y1 = fft(SpikeTrain(1,:),NFFT)/L1;
f1 = samplingrate/2*linspace(0,1,NFFT/2+1);

% Plot single-sided amplitude spectrum from raster plot corrupted with random
noise
figure(4)
plot(f1,2*abs(Y1(1:NFFT/2+1)))
title('Single-Sided Amplitude Spectrum of y(t)')
xlabel('Frequency (Hz)')
ylabel('|Y(f)|')
xlim([0 50])
```